\documentclass[a4paper,12pt]{amsart}
\usepackage{natbib}
\usepackage[T1]{fontenc}
\usepackage{amsmath}
\usepackage[obeyspaces]{url}
\usepackage{soul}

\title[On the possibility of unifying $\dots$]{On the possibility of unifying the electromagnetic and the gravitational fields}

\thanks{Originally published in German as ''\"Uber die M\"oglichkeit, das elektromagnetische Feld und das Gravitationsfeld zu vereinigen'', \textit{Physik. Zeitschr.} XV (1914) 504-506. Translated by Frank Borg, present address: University of Jyv\"askyl\"a, Chydenius Institute, POB 567, 67101-Karleby, Finland; email: \url{borgbros@netti.fi}. For an earlier translation by P.\ Freund, and a reprint of the original, see  T.\ Appelquist, A.\ Chodos, P.\ G.\ O.\ Freund (eds), \textit{Modern Kaluza-Klein theories}, Addison-Wesley 1987. A scanned pdf-version of the original paper along with the two subsequent papers that Nordstr{\"o}m published on the topic can be found at this url: \url{www.netti.fi/~borgbros/nordstrom}. For some brief historical sketches see Ch.\ Cronstr{\"o}m, ''Gunnar Nordstr{\"o}m (1881--1923)'' p.\ 3--9; S.\ Deser, ''The many dimensions of dimension'', p.\ 65--74; F.\ Ravndal, ''Scalar gravitation and extra dimensions'', p.\ 151--164: in Ch.\ Cronstr{\"o}m and C.\ Montonen (eds), \textit{Proceedings of the Gunnar  Nordstr{\"o}m Symposium on Theoretical Physics}, August 27--30, 2003, Helsinki. Commentationes Physico-Mathematicae 166/2004, The Finnish Society of Sciences and Letters.} 
\author{Gunnar Nordstr{\"o}m}

\begin{document}

\begin{abstract}
This is the first paper by Gunnar Nordstr{\"o}m (1881--1923) on his five dimensional theory. In his summary he states: ''It is shown, that a unifying treatment of the electromagnetic and gravitational fields is possible, if one considers the four dimensional spacetime world to be a surface in a five dimensional world.'' This paper was followed by two other papers written by Gunnar Nordstr{\"o}m during the one year period 1914--1915, on the same subject, which are also made available here. What Nordstr{\"o}m called ''Weltfl{\"a}che'' (world-surface) is nowadays termed a brane. Nordstr{\"o}m's theory was thus a precursor of the modern brane- and 5D-theories. His basic idea was to unify electromagnetism and gravitation (in the form of Nordstr{\"o}m's scalar theory) by extending the basic formalism of Maxwell's electromagnetic theory, as developed by Minkowski, to a five dimensional space of which physical spacetime is a hypersurface. Modern 5D-theories use the same basic approach but start instead from Einstein's general theory of relativity which is extended to the 5D-case. Einstein's theory was of course only in its final gestation when Nordstr{\"o}m worked on his theory, and the full impact of geometry on physics was only to become apparent later. Subsequently Theodor Kaluza (1921) and Oskar Klein (1926) also seized on the idea of spacetime extension as a path toward unification. -- F.B.

\end{abstract}

\maketitle


It is one of the great merits of the theory of relativity that it is able to represent the electromagnetic state of the aether using a vector, the \so{Minkowski} six-vector $\mathfrak{f}$, while the old formulation required two field vectors to represent this state. However, this possibility of characerizing the aether state becomes insufficient when one assumes, besides the electromagnetic field, a gravitational field. In the theories of gravitation, developed by \so{Mie}\footnote{G. \so{Mie}, Ann. d. Phys. \textbf{40}, 25, 1913.} and myself\footnote{G. \so{Nordstr{\"o}m}, diese Zeitschr. \textbf{13}, 1126, 1912; Ann. d. Phys. \textbf{40}, 872, 1913; \textbf{42}, 533, 1913.}, the gravitational field becomes a four-vector; if such a theory corresponds to reality, then the state of the aether will be characterized by a six-vector and a four-vector.

We are going the denote the components of the electromagnetic six-vector
by

\begin{equation*}
\mathfrak{f}_{xy}, \, \mathfrak{f}_{yz}, \, \mathfrak{f}_{zx}, \, \mathfrak{f}_{xu}, \, \mathfrak{f}_{yu}, \, \mathfrak{f}_{zu},
\end{equation*}

in which we have put $u = i c t$, where $c$ is the velocity of light. The components of the magnetic field strength $\mathfrak{H}$ and the electric field strength $\mathfrak{F}$ become then\footnote{H. \so{Minkowski}, G\"ott.\ Nachr.\ 1908, S.\ 58.}

\begin{equation}
\left\{
\begin{aligned}
\label{EQ:1}
\mathfrak{H}_x &= \mathfrak{f}_{yz} = -\mathfrak{f}_{zy} \quad \mbox{etc},\\
-i \mathfrak{F}_x &= \mathfrak{f}_{xu} = - \mathfrak{f}_{ux} \quad \mbox{etc}.
\end{aligned}
\right.
\end{equation}

We introduce next purely formally, for the components of the gravitational four-vector, the following symbols,

\begin{equation*}
\mathfrak{f}_{wx}, \, \mathfrak{f}_{wy}, \, \mathfrak{f}_{wz}, \, \mathfrak{f}_{wu}, 
\end{equation*}

(with $\mathfrak{f}_{xw} = - \mathfrak{f}_{wx}$, etc) and write the following system of equations:

{
\allowdisplaybreaks
\begin{align}
\label{EQ:I}
\nonumber
\frac{\partial \mathfrak{f}_{xy}}{\partial y} +
\frac{\partial \mathfrak{f}_{xz}}{\partial z} +
\frac{\partial \mathfrak{f}_{xu}}{\partial u} +
\frac{\partial \mathfrak{f}_{xw}}{\partial w} = \frac{1}{c} \mathfrak{k}_x ,\\
\nonumber
\frac{\partial \mathfrak{f}_{yx}}{\partial x} +
\frac{\partial \mathfrak{f}_{yz}}{\partial z} +
\frac{\partial \mathfrak{f}_{yu}}{\partial u} +
\frac{\partial \mathfrak{f}_{yw}}{\partial w} = \frac{1}{c} \mathfrak{k}_y ,\\
\tag{I}
\frac{\partial \mathfrak{f}_{zx}}{\partial x} +
\frac{\partial \mathfrak{f}_{zy}}{\partial y} +
\frac{\partial \mathfrak{f}_{zu}}{\partial u} +
\frac{\partial \mathfrak{f}_{zw}}{\partial w} = \frac{1}{c} \mathfrak{k}_z ,\\
\nonumber
\frac{\partial \mathfrak{f}_{ux}}{\partial x} +
\frac{\partial \mathfrak{f}_{uy}}{\partial y} +
\frac{\partial \mathfrak{f}_{uz}}{\partial z} +
\frac{\partial \mathfrak{f}_{uw}}{\partial w} = \frac{1}{c} \mathfrak{k}_u ,\\
\nonumber
\frac{\partial \mathfrak{f}_{wx}}{\partial x} +
\frac{\partial \mathfrak{f}_{wy}}{\partial y} +
\frac{\partial \mathfrak{f}_{wz}}{\partial z} +
\frac{\partial \mathfrak{f}_{wu}}{\partial u} = \frac{1}{c} \mathfrak{k}_w ,
\end{align}

\begin{align}
\label{EQ:II}
\nonumber
\frac{\partial \mathfrak{f}_{yz}}{\partial x} +
\frac{\partial \mathfrak{f}_{zx}}{\partial y} +
\frac{\partial \mathfrak{f}_{xy}}{\partial z} = 0 ,\\
\nonumber
\frac{\partial \mathfrak{f}_{zu}}{\partial y} +
\frac{\partial \mathfrak{f}_{uy}}{\partial z} +
\frac{\partial \mathfrak{f}_{yz}}{\partial u} = 0 ,\\
\nonumber
\frac{\partial \mathfrak{f}_{xu}}{\partial z} +
\frac{\partial \mathfrak{f}_{uz}}{\partial x} +
\frac{\partial \mathfrak{f}_{zx}}{\partial u} = 0 ,\\
\nonumber
\frac{\partial \mathfrak{f}_{yu}}{\partial x} +
\frac{\partial \mathfrak{f}_{ux}}{\partial y} +
\frac{\partial \mathfrak{f}_{xy}}{\partial u} = 0 ,\\
\nonumber
\frac{\partial \mathfrak{f}_{zw}}{\partial y} +
\frac{\partial \mathfrak{f}_{wy}}{\partial z} +
\frac{\partial \mathfrak{f}_{yz}}{\partial w} = 0 ,\\
\tag{II}
\frac{\partial \mathfrak{f}_{xw}}{\partial z} +
\frac{\partial \mathfrak{f}_{wz}}{\partial x} +
\frac{\partial \mathfrak{f}_{zx}}{\partial w} = 0 ,\\
\nonumber
\frac{\partial \mathfrak{f}_{yw}}{\partial x} +
\frac{\partial \mathfrak{f}_{wx}}{\partial y} +
\frac{\partial \mathfrak{f}_{xy}}{\partial w} = 0 ,\\
\nonumber
\frac{\partial \mathfrak{f}_{uw}}{\partial x} +
\frac{\partial \mathfrak{f}_{wx}}{\partial u} +
\frac{\partial \mathfrak{f}_{xu}}{\partial w} = 0 ,\\
\nonumber
\frac{\partial \mathfrak{f}_{uw}}{\partial y} +
\frac{\partial \mathfrak{f}_{wy}}{\partial u} +
\frac{\partial \mathfrak{f}_{yu}}{\partial w} = 0 ,\\
\nonumber
\frac{\partial \mathfrak{f}_{uw}}{\partial z} +
\frac{\partial \mathfrak{f}_{wz}}{\partial u} +
\frac{\partial \mathfrak{f}_{zu}}{\partial w} = 0 .
\end{align}
}

Both systems of equations are completely symmetric with respect to $x, y, z, u, w$. They have, of course, thus far no physical content; however, if we set all the partial derivatives with respect to $w$ equal to zero, then one finds that they transform into field equations of the electromagnetic and gravitational fields, when $\mathfrak{k}_x , \mathfrak{k}_y , \mathfrak{k}_z , \mathfrak{k}_u $ are the components of the four-current and $- \frac{1}{c} \mathfrak{k}_w$ is the rest mass-density of the gravitational mass.\footnote{ \label{FN:504} By this I understand the quantity, which I have denoted by $g \cdot \nu$ in the cited papers; thus,

\[
- \frac{1}{c} \mathfrak{k}_w = g \cdot \nu.
\]

For the components of the gravitational vector one might, more generally, introduce the representation $a \mathfrak{f}_{wx} , a \mathfrak{f}_{wy} , a  \mathfrak{f}_{wz} , a \mathfrak{f}_{wu}$ , where $a$ is an arbitrary real or imaginary constant; then we would have $ - \frac{a}{c} \mathfrak{k}_w = g \nu $. It follows, however, by applying the energy-momentum theorem, that $a$ must be either equal to +1 or -1, in order that the last equation in (\ref{EQ:I}) actually will be satisfied like the rest. 
} 

The four first equations in both systems become now the \so{Maxwell} equations in the form presented by \so{Minkowski}; the last equation in (\ref{EQ:I}) is the fundamental equation of gravitation, while the six remaining equations in (\ref{EQ:II}) express that the gravitational vector is irrotational.

This interpretation of the equations (\ref{EQ:I}), (\ref{EQ:II}) shows, that it is justified to consider the four dimensional spacetime as a surface situated in a five dimensional world. In such a five dimensional world the $\mathfrak{k}_m$ are the components of a five-vector and the $\mathfrak{f}_{mn}$ the components of a ten-vector; the latter gives a complete characterisation of the state of the aether. The five dimensional world has a distinguished axis, the $w$-axis; at every point of the four dimensional spacetime it is orthogonal to this axis and the derivatives of all the components of $\mathfrak{f}_{mn}$ with respect to $w$ are equal to zero. 

The components of $\mathfrak{f}$ can be expressed using a five-potential $\Phi_x$, $\Phi_y$, $\Phi_z$, $\Phi_u$, $\Phi_w$, when for each component

\begin{equation}
\label{EQ:2}
\mathfrak{f}_{mn} = \frac{\partial \Phi_n}{\partial m} -
\frac{\partial \Phi_m}{\partial n} .
\end{equation}

By differentiating the equations (\ref{EQ:I}) we obtain for the ''five-current'' $\mathfrak{k}$ the relation

\begin{equation}
\label{EQ:3}
\frac{\partial \mathfrak{k}_x}{\partial x} +
\frac{\partial \mathfrak{k}_y}{\partial y} +
\frac{\partial \mathfrak{k}_z}{\partial z} +
\frac{\partial \mathfrak{k}_u}{\partial u} +
\frac{\partial \mathfrak{k}_w}{\partial w} = 0,
\end{equation}

whence one may write for the five-potential the following six partial differential equations:

\begin{equation}
\label{EQ:4}
\left\{
\begin{aligned}
\frac{\partial \Phi_x}{\partial x} +
\frac{\partial \Phi_y}{\partial y} +
\frac{\partial \Phi_z}{\partial z} +
\frac{\partial \Phi_u}{\partial u} +
\frac{\partial \Phi_w}{\partial w} = 0,
\end{aligned}
\right.
\end{equation}

\begin{equation}
\label{EQ:5}
\left\{
\begin{aligned}
\frac{\partial^2 \Phi_x}{\partial x^2} +
\frac{\partial^2 \Phi_x}{\partial y^2} +
\frac{\partial^2 \Phi_x}{\partial x^2} +
\frac{\partial^2 \Phi_x}{\partial u^2} +
\frac{\partial^2 \Phi_x}{\partial w^2} &= - \frac{1}{c} \mathfrak{k}_x ,
\\
\dots &
\\
\frac{\partial^2 \Phi_w}{\partial x^2} +
\frac{\partial^2 \Phi_w}{\partial y^2} +
\frac{\partial^2 \Phi_w}{\partial x^2} +
\frac{\partial^2 \Phi_w}{\partial u^2} +
\frac{\partial^2 \Phi_w}{\partial w^2} &= - \frac{1}{c} \mathfrak{k}_w.
\end{aligned}
\right.
\end{equation}

In order to save space we have only explicitly written down two of the five equations in (\ref{EQ:5}). When these six conditions are satisfied, then the expressions (\ref{EQ:2}) will satisfy the field equations (\ref{EQ:I}), (\ref{EQ:II}) identically.

Our equations (\ref{EQ:I}), (\ref{EQ:II}) make it possible to give a unified presentation of the energy-momentum theorem for the combined electromagnetic and gravitational field. In order to obtain this theorem for the $x$-direction, 
we have to multiply each of the four equations (\ref{EQ:I}), which refer to the other coordinate-axes, with $\mathfrak{f}_{xy}$, $\mathfrak{f}_{xz}$, $\mathfrak{f}_{xu}$, $\mathfrak{f}_{xw}$, and each of the six equations (\ref{EQ:II}) which contain $x$, with $\mathfrak{f}_{mn}$, where $m$, $n$  denote the other two indices appearing in the equation in correctly ordered. The ten equations thus obtained are then added together after which a simple rearranging gives the desired result.

We will demonstrate the theorem for the $u$-direction -- that is, the energy theorem -- and must thus multiply the three first equations in (\ref{EQ:I}) with $\mathfrak{f}_{ux}$, $\mathfrak{f}_{uy}$, $\mathfrak{f}_{uz}$, and the last one with $\mathfrak{f}_{uw}$. Furthermore, one must multiply the second, third, fourth, eighth, ninth and tenth equation in (\ref{EQ:II}) with $\mathfrak{f}_{yz}$, $\mathfrak{f}_{zx}$, $\mathfrak{f}_{xy}$, $\mathfrak{f}_{wx}$, $\mathfrak{f}_{wy}$, $\mathfrak{f}_{wz}$, respectively. We obtain, after addition and some rearranging, the term

{
\allowdisplaybreaks
\begin{align*}
\mathfrak{f}_{ux} 
\left( 
\frac{\partial \mathfrak{f}_{xy}}{\partial y} +
\frac{\partial \mathfrak{f}_{xz}}{\partial z} +
\frac{\partial \mathfrak{f}_{xw}}{\partial w}
\right) + 
\\
\mathfrak{f}_{uy} 
\left( 
\frac{\partial \mathfrak{f}_{yx}}{\partial x} +
\frac{\partial \mathfrak{f}_{yz}}{\partial z} +
\frac{\partial \mathfrak{f}_{yw}}{\partial w}
\right) + 
\\
\mathfrak{f}_{uz} 
\left( 
\frac{\partial \mathfrak{f}_{zx}}{\partial x} +
\frac{\partial \mathfrak{f}_{zy}}{\partial y} +
\frac{\partial \mathfrak{f}_{zw}}{\partial w}
\right) + 
\\
\mathfrak{f}_{uw} 
\left( 
\frac{\partial \mathfrak{f}_{wx}}{\partial x} +
\frac{\partial \mathfrak{f}_{wy}}{\partial y} +
\frac{\partial \mathfrak{f}_{wz}}{\partial z}
\right) + 
\\
\mathfrak{f}_{yz} 
\left( 
\frac{\partial \mathfrak{f}_{zy}}{\partial u} +
\frac{\partial \mathfrak{f}_{uy}}{\partial z} 
\right) + 
\mathfrak{f}_{zx}
\left( 
\frac{\partial \mathfrak{f}_{xu}}{\partial z} +
\frac{\partial \mathfrak{f}_{uz}}{\partial x} 
\right) + 
\\
\mathfrak{f}_{xy} 
\left( 
\frac{\partial \mathfrak{f}_{yu}}{\partial x} +
\frac{\partial \mathfrak{f}_{ux}}{\partial y} 
\right) + 
\mathfrak{f}_{wx}
\left( 
\frac{\partial \mathfrak{f}_{uw}}{\partial x} +
\frac{\partial \mathfrak{f}_{xu}}{\partial w} 
\right) + 
\\
\mathfrak{f}_{wy}
\left( 
\frac{\partial \mathfrak{f}_{uw}}{\partial y} +
\frac{\partial \mathfrak{f}_{yu}}{\partial w} 
\right) + 
\mathfrak{f}_{wz}
\left( 
\frac{\partial \mathfrak{f}_{uw}}{\partial z} +
\frac{\partial \mathfrak{f}_{zu}}{\partial w} 
\right) + 
\\
\mathfrak{f}_{ux} \frac{\partial \mathfrak{f}_{xu}}{\partial u} +
\mathfrak{f}_{uy} \frac{\partial \mathfrak{f}_{yu}}{\partial u} +
\mathfrak{f}_{uz} \frac{\partial \mathfrak{f}_{zu}}{\partial u} +
\mathfrak{f}_{uw} \frac{\partial \mathfrak{f}_{wu}}{\partial u} +
\\
\mathfrak{f}_{yz} \frac{\partial \mathfrak{f}_{yz}}{\partial u} +
\mathfrak{f}_{zx} \frac{\partial \mathfrak{f}_{zx}}{\partial u} +
\mathfrak{f}_{xy} \frac{\partial \mathfrak{f}_{xy}}{\partial u} +
\\
\mathfrak{f}_{wx} \frac{\partial \mathfrak{f}_{wx}}{\partial u} +
\mathfrak{f}_{wy} \frac{\partial \mathfrak{f}_{wy}}{\partial u} +
\mathfrak{f}_{wz} \frac{\partial \mathfrak{f}_{wz}}{\partial u} +
=
\\
\frac{1}{c} 
\left(
\mathfrak{f}_{ux} \mathfrak{k}_x +
\mathfrak{f}_{uy} \mathfrak{k}_y +
\mathfrak{f}_{uz} \mathfrak{k}_z +
\mathfrak{f}_{uw} \mathfrak{k}_w
\right) .
\end{align*}
}
 
A simple rearranging gives the desired equation

\begin{align}
\label{EQ:6}
\frac{\partial}{\partial x} 
\left(
\mathfrak{f}_{uy} \mathfrak{f}_{yx} +
\mathfrak{f}_{uz} \mathfrak{f}_{zx} +
\mathfrak{f}_{uw} \mathfrak{f}_{wx} 
\right) + 
\\
\nonumber
\frac{\partial}{\partial y} 
\left(
\mathfrak{f}_{ux} \mathfrak{f}_{xy} + \dots
\right) +
\frac{\partial}{\partial z} 
\left(
\mathfrak{f}_{ux} \mathfrak{f}_{xz} + \dots
\right) +
\frac{\partial}{\partial w} 
\left(
\mathfrak{f}_{ux} \mathfrak{f}_{xw} + \dots
\right) + 
\\
\nonumber
\frac{1}{2} \frac{\partial}{\partial w}
\left(
- \mathfrak{f}_{ux}^2 -
\mathfrak{f}_{uy}^2 - \mathfrak{f}_{uz}^2 - \mathfrak{f}_{uw}^2 + \mathfrak{f}_{yz}^2 + \mathfrak{f}_{zx}^2 + 
\mathfrak{f}_{xy}^2 + \mathfrak{f}_{wx}^2 + \mathfrak{f}_{wy}^2 + \mathfrak{f}_{wz}^2
\right) =
\\
\nonumber
\frac{1}{c} 
\left(
\mathfrak{f}_{ux} \mathfrak{k}_x +
\mathfrak{f}_{uy} \mathfrak{k}_y +
\mathfrak{f}_{uz} \mathfrak{k}_z +
\mathfrak{f}_{uw} \mathfrak{k}_w
\right) .
\end{align}

The quantities in the brackets on the left hand side are components of a five dimensional tensor, and the equation is thus of the form

\begin{equation}
\label{EQ:6a}
\tag{6 a}
 \frac{\partial P_{ux}}{\partial x} +
 \frac{\partial P_{uy}}{\partial y} +
 \frac{\partial P_{uz}}{\partial z} +
 \frac{\partial P_{uu}}{\partial u} +
 \frac{\partial P_{uw}}{\partial w} = \mathfrak{K}_u .
 \end{equation}

If we multiply it with $i c$ and set $\frac{\partial P_{uw}}{\partial w}$ equal to zero we obtain the energy relation in its conventional form. In this way we obtain for the $x$-component of the energy-current

\begin{align}
\label{EQ:7}
\mathfrak{S}_x = ic
\left(
\mathfrak{f}_{uy} \mathfrak{f}_{yx} +
\mathfrak{f}_{uz} \mathfrak{f}_{zx} +
\mathfrak{f}_{uw} \mathfrak{f}_{wx}
\right) =
\\
\nonumber
c 
\left(                       
\mathfrak{F}_y \mathfrak{H}_z - \mathfrak{F}_z \mathfrak{H}_y  
\right) -
\frac{\partial \Phi_w}{\partial t} \frac{\partial \Phi_w}{\partial x}.
\end{align}

The last expression is obtained by using the equations (\ref{EQ:1}) and
(\ref{EQ:2}), where the derivatives with respect to $w$ are set equal to zero. Using vector analysis the expression for $\mathfrak{S}$ can be rendered as

\begin{equation}
\label{EQ:7a}
\tag{7 a}
\mathfrak{S} = c [\mathfrak{F} \, \mathfrak{H}] - \frac{\partial \Phi_w}{\partial t} \, \nabla \Phi_w .
\end{equation}

For the energy-density we obtain likewise from (\ref{EQ:6}) a vector analytical form:

\begin{equation}
\label{EQ:8}
\psi = \frac{1}{2} 
\left\{
\mathfrak{F}^2 + \mathfrak{H}^2 + \left( \nabla \Phi_w \right)^2 + 
\frac{1}{c^2} \left( \frac{\partial \Phi_w}{\partial t} \right)^2 
\right\} .
\end{equation}

These expressions for the energy-current and the energy-density are the sums of their familiar expressions in case of the the electromagnetic field and the gravitational field; which is exactly the desired result of our considerations.

It is easy to see that the claim made in the note on p.\ 504 [p.~\pageref{FN:504}] left column is correct, and that only the components of the ten-vector $\mathfrak{f}$ and
the five-vector $\mathfrak{k}$ having the index $u$ are imaginary.

By permutating the indices in equation (\ref{EQ:6}) one obtains four additional equations. The three, which refer to the spatial directions, naturally express the momentum theorem in the familiar form. The equation for the $w$-direction becomes on vector analytical form, after introducing the field strengths and the gravitational potential $\Phi_w$ 

\begin{align*}
\label{EQ:wequ}
- \, \mbox{div} \left\{
[\mathfrak{H} , \nabla \Phi_w] + \frac{1}{c} \mathfrak{F} \frac{\partial \Phi_w}{\partial t}
\right\} +
\frac{1}{c} \frac{\partial}{\partial t} 
\left(
\mathfrak{F} \nabla \Phi_w 
\right) + 
\\
\frac{1}{2} \frac{\partial}{\partial w}
\left\{
\mathfrak{H}^2 - \mathfrak{F}^2 - \left( \nabla \Phi_w \right)^2 +
\frac{1}{c^2} 
\left( \frac{\partial \Phi_w}{\partial t} \right)^2
\right\} =
\\
- \frac{1}{c}
\left\{
\mathfrak{k} \nabla \Phi_w + \mathfrak{k}_u \frac{\partial \Phi_w}{\partial u} 
\right\} .
\end{align*}

Presently it is not known whether this expression might have a physical meaning.

The above point of view provides, as we have seen, some formal advantages as it permits the electromagnetic and the gravitational fields to be expressed as a single field. Of course, the equations have not thereby gained any new physical meaning. Yet I do not think it can be excluded that this symmetry may have a deeper foundation. However, I will not here explore the possibilities that one might conjecture regarding this.   

\so{Summary}. It is shown that a unifying treatment of the electromagnetic and gravitational fields is possible if one considers the four dimensional spacetime-world to be a surface in a five dimensional world.
\\[10pt] {\hspace*{20pt}\so{Helsingfors}, 30. March 1914.}\\[15pt] \rightline{{\footnotesize (Submitted 3. April 1914)}}
\\[20pt]
\centerline{\rule[2mm]{5cm}{0.5mm}}
\\

\appendix

\section{A brief summary using modern notation}

Some comments using a modernized notation is offered here for a quick review of the main contents of the paper. (Because of the $u = ict$ convention the metric signature here is (+ + + + + ).) We use $A$ for the ''vector potential'' generalized to five dimensions, and $F$ for the corresponding generalized field ($i = 1, \dots ,5$, with $i$ = 5 corresponding to $w$ in the paper),

\begin{align*}
&A = A_i dx^i,
\\
&F = dA = \frac{1}{2}\left( \partial^i A_j 
- \partial^j A_i \right) dx^i \wedge dx^j. 
\end{align*}

The equations (\ref{EQ:II}) are thus the same as $dF = 0$. Using the convention of summing over repeated indices, the equations (\ref{EQ:I}) can be written as

\[
\partial_j F^{ij} = J^i,
\]

where we have used $J$ for the generalized current density. The current density may be represented as a 4-form given by

\[
J = J^i \epsilon_{ijklm} dx^j \wedge dx^k \wedge dx^l \wedge dx^m, 
\]

where $\epsilon_{ijklm}$ is the anti-symmetric tensor with values $\pm 1$. By defining the dual $^\star F$ of $F$ as

\[
^\star F = \frac{1}{2} \epsilon^{ijklm} F_{lm} dx^i \wedge dx^j \wedge dx^k,
\]

then the equations (\ref{EQ:I}) can be written as

\[
d\, {^\star F} = J.
\]

The equation (\ref{EQ:4}) is a generalization of the ''Lorentz gauge condition'', $\partial_i A^i = 0$, to five dimensions which together with previous equation gives

\[
\partial_i \partial^i A^j = -J^j,
\]

which is equation (\ref{EQ:5}).

The key observation of the paper is that the ansatz $\partial^5 A^k = 0$ splits the previous equations into the ''wave equations'' for electromagnetism and scalar gravity ($A^5$ is associated with the gravitational potential and $J^5$ with the matter density). From $d\, {^\star F} = J$ we immediately get $dJ = 0$ which states the conservation of the five-current, equation (\ref{EQ:3}). Equation $dF = 0$ is equvialent with

\[
\sum_{\text{cycl}} \partial_k F_{ij} = 0.
\]

Using this and the field equation $\partial_j F^{ij} = J^i$ we can demonstrate equation (\ref{EQ:6a}) in the form

\[
\partial_j T^{ij} = F^{ij} J_j
\]

with the ''energy-momentum tensor'' $T^{ij}$ defined by\footnote{The general concept of the energy-momentum tensor was codified by Max von Laue, Ann.\ d.\ Phys.\ \textbf{35}: 524--542, 1911; Das Relativit\"atsprinzip, Braunschweig 1911, which Nordstr\"om had studied closely. For the historical background see J D Norton, ''Einstein, Nordstr\"om and the early demise of the scalar Lorentz covariant theories of gravitations,'' Archive for History of Exact Sciences, \textbf{45}: 17--94, 1992; \url{www.pitt.edu/~jdnorton}.}

\[
T^{ij} = F^i_{~ k} F^{kj} + \frac{1}{4} \delta^{ij} F_{kl} F^{kl}.
\]

While $\partial_j T^{ij} = 0$ for $i = 1, 2, 3, 4$ is associated with momentum and energy conservation, Nordstr{\"o}m is not sure what to make of the case $i = 5$. It was only in 1918 that Emmy Noether\footnote{Nachr.\ d.\ K\"onig.\ Gesellsch.\ d.\ Wiss.\ zu G\"ottingen, Math-phys.\ Klasse, 235--257, 1918; for an engl.\ translation see \url{physics/0503066}.} published her theorem on the relation between symmetry and conserved quantities for theories defined in terms of Lagrangians. In this case, the Lagrangian density (inclusive the interaction term) is given by

\[
\mathcal{L} = -\frac{1}{4} F_{ij} F^{ij} + A_i J^i. 
\]

Thus, with Noether's methods we can derive the energy-momentum tensor from the Lagrangian by considering translations along the $x^i$-directions. The corresponding conserved ''charges'' are defined by integrating over hypersurfaces $t$ = constant. Define the ''surface'' forms by

\[
dS_i = (-1)^{i+1} dx^1 \wedge \cdots \widehat{dx{^i}} \cdots \wedge dx^5
\]

where the ''widehat'' means that the corresponding element is absent. Then if we define $V_t$ to be the hypersurface $x^4 = i c t$ it follows that

\begin{align*}   
&\int_{V_{t_1}} T^{ij} dS_j - \int_{V_{t_0}} T^{ij} dS_j = 
\int_{\partial W} T^{ij} dS_j = 
\\
&\int_{W} d(T^{ij} dS_j) =
\int_{W} \partial_j T^{ij} d\Omega = 0,
\end{align*}

where $W$ is the 5-volume bounded by $V_{t_0}$, $V_{t_1}$, and $d\Omega$ is the 5-volume form. (We note that if the fields are constant along the $x^5$-dimension, according to the ansatz, then the $x^5$-integration only adds a spurious factor.) Thus $P^i \equiv \int_{V_{t}} T^{ij} dS_j$ is independent of the time slice, which is the content of the conservation theorem. The  conserved quantity $P^5$ associated with $x^5$-translation would then correspond to the integral $\int T^{54} dx dy dz$ = $-\int \mathbf{E} \cdot \nabla A^5 dx dy dz$. What is lacking here is the corresponding expression for the matter tensor $T_m$ and the ''5-momentum'' of a particle which would balance the field contribution such that $\partial_j T_{\text{tot}}^{ij} = 0$.

\end{document}